\documentclass[twocolumn,prb, showpacs,showkeys]{revtex4-1}
\usepackage{hyperref}
\usepackage[dvipdfm]{graphicx}
\usepackage{amsmath}
\usepackage{amssymb}
\usepackage{amsfonts}
\usepackage{booktabs}    
\usepackage{subfigure}
\usepackage{longtable}
\usepackage{color}

\begin{document}
\title{One-dimensional physics in transition-metal nanowires:\\
Phases and elementary excitations }
\author{Jun-ichi Okamoto}
\email{okamoto@phys.columbia.edu}
\affiliation{Department of Physics, Columbia University, 538 West 120th Street, New York, New York 10027, USA}

\author{A. J. Millis}
\email{millis@phys.columbia.edu}
\affiliation{Department of Physics, Columbia University, 538 West 120th Street, New York, New York 10027, USA}

\date{\today}

\begin{abstract}
We used the Hartree-Fock approximation to classify the electronic phases that might occur in a transition metal nanowire. The important features of this situation are orbital degeneracy (or near-degeneracy) and interactions favoring locally high-spin configurations.  In this circumstance, spin density wave and triplet superconductivity states are favored. If the interactions favor locally low-spin configurations as in the previously studied spin ladder systems, orbital density wave and singlet superconductivity are observed. 
\end{abstract}


\maketitle
\section{Introduction}
Self-assembly epitaxial techniques have enabled the fabrication of one-dimensional (1D) atomic wires composed of adatoms confined at step edges on surfaces of appropriately chosen substrates \cite{Snijders2010, Barth2005, Himpsel2001}. In many physically relevant cases, the surface bandgap structure of the substrate material is such that the electronic states of the adatoms are decoupled from the bulk substrate bands (at least to leading order) and a one-dimensional electron system can be realized.  STM (Scanning Tunneling Microscope) and ARPES (Angle-Resolved Photoemission) measurements of Au nanowires grown on the Si(577) surface  show that a charge-density wave (CDW) occurs at low temperatures\cite{Ahn2003, Yeom2005}. The physics of Au nanowires is still the subject of debate\cite{Snijders2010, Erwin2010}, but it appears that the relevant band is derived from Au $s$-states for which electron-electron interactions are relatively weak, and the dominant physics may be associated with lattice instabilities\cite{Johannes2008}. However, many other adatom/substrate pairs are possible, and this opens new possibilities including the study of one-dimensional electron gases formed from transition metal $d$-orbitals. For example, wires composed of Co adatoms have been furrowed on a Cu substrate\cite{Wang2008,Zaki2009}.

From the theoretical point of view, the important features of transition metal-based wires are  the orbital degeneracy of the transition-metal $d$-levels, which permits a rich set of on-site interactions and the small size of the orbitals, which leads to larger interaction effects.  In particular, the Hund coupling favors locally high spin configurations, potentially leading to interesting spin structures. Motivated by these ideas and the recent experimental success \cite{Wang2008,Zaki2009}, in this paper, we investigate the physics of a nanowire in which the important electronic states are derived from the transition metal $d$-orbitals. We use mean-field theory to establish the phase diagram and elucidate the general classes of behavior.  While mean-field theory is not an exact description of interacting electron systems especially in low spatial dimensions, it should tell reasonable indication of what physics is relevant and provide a starting point for more exact treatment. A subsequent paper will use renormalization group and bosonization methods to obtain a detailed picture of the same model when one-dimensional nature is significant. 

Consideration of the physics of transition metals leads to models with multiple electronic bands with more or less arbitrary interactions. Models of this general class have been previously considered in the literature, both for their intrinsic interest \cite{Schulz1996}, and as steps toward understanding heavy fermion \cite{Varma1985a, Strong1994a, Fujimoto1997, Shelton1996b} and high temperature superconductor systems \cite{Fabrizio1992,Finkelstein1993,Khveshchenko1994d,Balents1996,Tsuchiizu2002,Wu2003}. In these models the multiple bands arise from physically different atoms: in the heavy fermion case, one band represents the local moments and the other the wide band of conduction electrons; in the high $T_c$ case an important motivation has been models of ``spin-ladder'' compounds\cite{Rice1993}. While a formal mapping may be established between these models and the models of interest here, the different structure of the interactions leads to different physical behaviors. Models more directly analogous to those of present interest have also been investigated, and the focus has been put on ferromagnetism and orbital ordering in the strong coupling regime\cite{Roth1966, Penn1966, Kugel1972, Cyrot1975, Gill1987, Sakamoto2002}. The relation of these works to the results obtained here  is discussed below in Sections \ref{model}, \ref{method} and \ref{strong-coupling}.

The organization of this paper is as follows. In Sec. \ref{model} we explain the model and the symmetries. In Sec. \ref{order parameters}, we define order parameters encoding the physics arising from the backscattering and forward scattering. The general properties of mean-field solution and approximation we employed will be presented in Sec. \ref{method}. In section \ref{weak-coupling phase diagrams}, we show the obtained phase diagrams in the weak-coupling limit, and discuss the results. Sec. \ref{strong-coupling} is devoted to phase diagrams obtained in the strong coupling limit. Finally, Sec. \ref{summary} is conclusion and summary.

\section{Model}
\label{model}
\begin{figure}[!htb]
\centering
\subfigure[Two-leg Hubbard ladder]{
\includegraphics[scale=1]{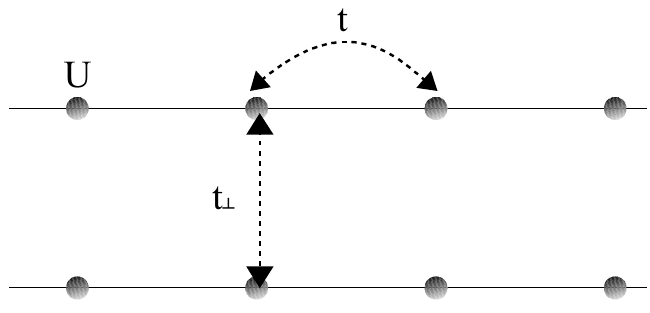}}
\subfigure[Atomic wire with $d$-orbitals]{
\includegraphics[scale=0.8]{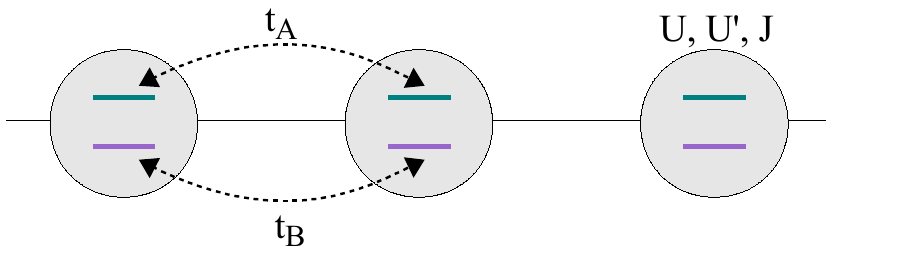}}
\caption{\label{wire} Schematic pictures of two-leg Hubbard ladders and $d$-orbital model. In two-leg Hubbard model, there is hopping along the ladder, $t$, and transverse hopping, $t_{\perp}$. The on-site interaction is $U$. For atomic wire with $d$-orbitals, we have intraband hopping, $t_{A}$ and $t_{B}$, with various on-site interactions, $U$, $U'$, and $J$.}
\end{figure}

While many of our considerations are general, we are motivated in particular by recent success growing monatomic Co chains at step edges on a Cu surface\cite{Wang2008,Zaki2009}. Band calculations indicate that the Co electrons are hybridized into the bulk Cu states mostly away from Fermi energy, and Co electrons form one-dimensional partly filled bands near Fermi energy\cite{Zaki}. Thus, as a first approximation, we may consider that the system is described by a multi-orbital Hubbard-like model representing the Co $d$-orbitals with local onsite Coulomb interactions
\begin{equation}
H = \sum_{\langle i , j \rangle } \sum _{m, s} -t^{m m'}_{i j}  \left( c^{\dagger}_{i m s}  c_{j m' s} + \text{H.c.} \right) + H_{\text{int}}.
\label{kin}
\end{equation}
Here $c^{(\dagger)}_{i m s}$ is the annihilation (creation) operator for a $d$-electron in orbital $m$ with spin $s$ at site $i$. $t_{ij}^{mm'}$ is the hopping between from orbital $m$ on site $i$ to orbital $m'$ on site $j$. The interaction terms $H_{\text{int}}$  will be discussed below. Through the paper, we set the  lattice constant equal to 1. The presence of the surface breaks the symmetry between $d$ levels and may lead to an arbitrary ionization level. For the sake of simplicity, we will consider here only the case where Fermi energy crosses two orbitals, $m=A, B$, although in the general case one would have $5$ $d$-derived bands with an arbitrary fermi energy. Furthermore, the rotational symmetry in $H_{int}$ as we will see always allows us to diagonalize the hopping matrix, so we will ignore $t^{AB}$. 

In the weak-coupling limit, the band structure is characterized by four Fermi points: two Fermi momentum, $k_A$ and $k_B$, and two chiralities, $r = R, L$. $R (L)$ represents electrons around positive (negative) Fermi momenta. The total particle number is  $n=2(k_A+k_B)/\pi$. It is useful to distinguish the two cases of half filling ($n=2$) and arbitrary filling ($n\neq 2$). For arbitrary filling we expect in general that  $k_A\neq k_B\neq \pi/2$ but two special cases are possible: we may have  degenerate bands,  $k_A=k_B$; alternatively, one of the two bands may be half-filled  i.e. $k_A = \frac{\pi}{2} \neq k_B$. If $n=2$ then $k_A + k_B = \pi$; in general we expect $k_A \neq k_B$ but both bands could be commensurate, $k_A = k_B = \frac{\pi}{2}$. When two bands have equal Fermi momentum and Fermi velocities, the kinetic term acquires $O(3)\simeq SU(2)$ orbital symmetry.

For the two-orbital system the interaction terms have the following form:
\begin{equation}
\begin{split}
H_{\text{int}} &= U  \sum_{i, m} n_{i m \uparrow} n_{i m \downarrow} \\
&+ U' \sum _{i,  s} n_{i A s} n_{i B \bar{s}}\\
&+ (U' -J) \sum _{i, s} n_{i A s} n_{i B s}\\
&- J \sum_{i, s} c^{\dagger}_{i A s}  c_{i A \bar{s}} c^{\dagger}_{i B \bar{s}} c_{i B s} \\
&+J' \sum_{i } \left( c^{\dagger}_{i A \uparrow} c^{\dagger}_{i A \downarrow} c_{i B \downarrow} c_{i B \uparrow}  + \text{H.c.} \right)
\label{int}
\end{split}
\end{equation}
where and $n_{i m s} = c^{\dagger}_{i m s} c_{i m s}$ is the electron density and $\bar{s} = -s$. $U$ and $U'$ indicates on-site Coulomb repulsion between two electron in the same band or different bands, and $J$ represents Hund coupling favoring high spin state. $J'$ is the so-called pair-hopping term. For a transition metal ion in free space all of these parameters are positive. We assume that the symmetry breaking by substrate primarily affects the hopping terms in the Hamiltonian without changing the local orbitals too much. This enables us to use free-space rotation symmetries to reduce the number of interaction constant \cite{Dagotto2000}. In this case we have 
\begin{gather}
J = J' \\
U = U' + 2J.
\label{coulomb integrals}
\end{gather}
The first equality is derived from the fact that Wannier wave functions are real, and the second one represents rotational invariance in orbital space. With this simplification, the interaction terms now have $U(1)$ orbital rotational symmetry about $y$-axis. For a transition metal ion in free space, $U\gg J>0$, so that all interaction parameters are positive. Screening will reduce the value of $U$, but will lead to only negligible changes in  $J$ \cite{Aryasetiawan2006}, and most calculations indicate that even the reduced value of $U$ is greater than $J$.

Now we talk about the symmetry of the Hamiltonian in Eq. \eqref{kin} and Eq. \eqref{int}. The model has $SU(2)$ symmetry about rotation of spins. The $U(1)$ rotational symmetry in orbital space in interaction terms is preserved only when two bands are equivalent. Otherwise, the orbital symmetry of the total Hamiltonian is completely broken.

Lastly we compare our models to previously studied ones. In particular, this model is similar to two-leg Hubbard ladder models, which have been studied extensively\cite{Varma1985a,Fabrizio1992,Finkelstein1993,Strong1994a,Khveshchenko1994d,Fujimoto1995,Balents1996,Schulz1996,Shelton1996b,Fujimoto1997,Lin1998,Azaria2000,Tsuchiizu2002,Wu2003,Lee2004,Controzzi2005, Shirakawa2008,Nonne2010,Tsvelik2011}. The comparison is sketched in Fig. \ref{wire}. In essence the two sides of a rung of the ladder (or, more precisely, the odd and even parity combinations of these states when there is strong transverse hopping) map to the two atomic states we consider. Generically, in the ladder problem, the hopping across a rung is non-vanishing, implying in our language  $k_{A} \neq k_{B}$. However, the interactions of two models are quite different. In two-leg ladder problems, the intra-chain interactions $U$ and transverse hopping $t_{\perp}$ are supposed to induce effective antiferromagnetic coupling ($- U/t_{\perp}^2 < 0$) between two sites connected by a rung of the ladder\cite{Fabrizio1992, Finkelstein1993, Khveshchenko1994d, Schulz1996, Balents1996}. On the other hand, the model considered here doesn't have such hopping since two states in an atom are orthogonal, but instead it has Hund coupling ($J>0$), which favors high-spin states. 

More recently several groups attempted a general classification of the physics of ladder systems with generic interactions either with transverse hopping \cite{Tsuchiizu2002, Wu2003}, or without transverse hopping\cite{Nonne2010}. When two Fermi momenta are different, $k_{A}\neq k_{B}$, our model is considered in Refs. \onlinecite{Tsuchiizu2002} and \onlinecite{Wu2003}, while when $k_{A}=k_{B}$, we have pair-hopping term which is not included in Ref. \onlinecite{Nonne2010}. These studies employ perturbative renormalization group and bosonization assuming weak interactions and one-dimensionality. While we will pursue this direction in subsequent paper, here in this paper we will use a mean-field approach, which allows us to access the strong coupling regime of the model -- which is hard to success by perturbative RG and bosonization -- as well as to classify the different possible states.

More directly related studies has been done on the same model focusing on ferromagnetism and orbital orders\cite{Roth1966, Penn1966, Kugel1972, Cyrot1975, Gill1987, Sakamoto2002}. In the strong coupling regime, $U\gg J\gg t$, both analytical and numerical calculation show ferromagnetism and orbital antiferromagnetism as a ground state at quarter-filling. Around quarter-filling, ferromagnetic state is robust to hole-doping and electron-doping with nearest neighbor hopping. However, below quarter-filling, the inclusion of further hopping leads to disappearance of ferromagnetism while the state still exists above quarter-filling. Haldane gapped state with $S=1$ is expected at half-filling.

\begin{table*}[!ht]
\begin{ruledtabular}
\centering
\begin{tabular}{c|c|cccc}
$(i,j)$&Particle-hole order & Particle-particle order & $P$ & $L$ & $S$\\ 
\hline
$(0,0)$ & Charge density wave (CDW) & $d'$-wave singlet SC($d'$SS) & -1 & 0 & 0\\
$(0,3)$& Spin density wave (SDW) &$p_{y}'$-wave triplet SC ($p_{y}'$TS) &  1 & 0 & 1 \\
$(1,0)$& $s'$-wave charge density wave ($s'$CDW)  & $p_{y}$-wave singlet SC ($p_{y}$SS) & 1 & 1 & 0\\
$(1,3)$& $s'$-wave spin density wave ($s'$SDW)  & $d$-wave triplet SC ($d$TS) & -1 & 1 & 1 \\
$(2,0)$ & $p_{y}'$-wave charge density wave ($p'$CDW)  &  $s$-wave singlet SC ($s$SS) & 1 & 1 &0\\
$(2,3)$ & $p_{y}'$-wave spin density wave ($p'$SDW) &  $p_{x}$-wave triplet SC ($p_{x}$TS) &-1 & 1 &1\\
$(3,0)$ & $p_{y}$-wave charge density wave ($p$CDW) &  $s'$-wave singlet SC ($s'$SS)& 1 & 1 &0\\
$(3,3)$ & $p_{y}$-wave spin density wave ($p$SDW) &$p_{x}'$-wave triplet SC ($p_{x}'$TS) &-1 & 1 &1
\end{tabular}
\caption{Classification of order parameters. `` ' '' indicates that the order is interband type. The eigenvalues of each superconducting phase under parity ($P$), orbital rotation ($L$), and spin rotation ($S$), are also listed. Here the orbital ($i$) and spin ($j$) indices are defined in Eq. \eqref{ph} and Eq. \eqref{pp}. Particle-hole channels are even under parity.}
\label{order}
\end{ruledtabular}
\end{table*}

\section{Order parameters}
\label{order parameters}
Mean-field theory involves minimizing the energy with respect to a free-fermion density matrix characterized by non-vanishing expectation values for fermion bilinears in the particle-hole (ph) and particle-particle (pp) channels. A general bilinears is characterized by a momentum, $q$, spin, $s$, and orbital, $m$. We distinguish between $q=0$ cases (ferromagnetism, ferro-orbital order, and superconductivity), and $q \neq 0$ (charge/spin density wave, orbital density wave, FFLO or pair-density wave superconducting states). We will mostly not be interested in the dependence of the expectation values on the magnitude of the fermion momentum and will mostly be interested in electronic states near the fermi points. Therefore, we label the bilinears by the chirality, spin, and orbital indices without explicitly denoting the $q$ or fermion momentum. The basic objects are:
\begin{equation}
(\Delta_{ph})_{rr'}^{ss';mm'}= \frac{1}{N} \sum_{k\sim k_F}\langle c^\dagger_{rms}c_{r'm's'}\rangle 
\label{ph}
\end{equation}
and particle-particle bilinears,
\begin{equation}
(\Delta_{pp})_{rr'}^{ss';mm'}= \frac{1}{N} \sum_{k\sim k_F}\langle m s c^\dagger_{rms}c^{\dagger}_{r'\overline{m'} \overline{s'}}\rangle
\label{pp}
\end{equation}
where $c^{(\dagger)}_{rms}$ is the annihilation (creation) operator of electron with chirality $r$, orbital $m$, and spin $s$. We will use the following convenient basis to represent these,
\begin{gather}
\mathcal{O}^{ij}_{ph} =\sum_{mm'ss'} \tau^{i}_{mm'} \sigma^{j}_{ss'} (\Delta_{ph})^{ss';mm'} + \text{H.c.}\\
\mathcal{O}^{ij}_{pp} = \sum_{mm'ss'} \tau^{i}_{mm'} \sigma^{j}_{ss'} (\Delta_{ph})^{ss';mm'} + \text{H.c.}
\end{gather}
where $i, j = (0, 1, 2, 3)$ and $\tau$ and $\sigma$ are Pauli matrices with $\tau^{0}_{ab} = \sigma^{0}_{ab} = \delta_{ab}$. These operators transforms as tensor of rank 2 under the rotation of $SO(4)\simeq SU(2)_{\text{spin}} \times SU(2)_{\text{orbital}}$. Due to the $SU(2)_\text{spin}$ symmetry of the Hamiltonian, we can take quantization axis along $z$-direction for spins, and consider only $j=0$ and 3. However, we will keep all $i=0\sim3$, since there are cases without any orbital  symmetry. $\sigma^{0(3)}$ combination gives spin singlet (triplet), that is, charge (spin) mode. Similarly $\tau^{0(3)}$ gives orbital singlet (triplet), and $\tau^{1(2)}$ gives symmetric (anti-symmetric) combinations of orbitals. 

We first discuss the cases with $r=r'$ in Eq. \eqref{ph}, which corresponds to spatially uniform density order. We consider only ferromagnetism (FM), orbital-ferromagnetism (OFM), and combinations of these two. These orders are characterized by non-zero density polarizations:
\begin{equation}
\begin{split}
\Delta_{FM} &= \langle n_{A\uparrow}\rangle   - \langle n_{A\downarrow}\rangle  + \langle n_{B\uparrow}\rangle  -\langle n_{B\downarrow}\rangle \neq 0\\
\Delta_{OFM} &= \langle n_{A\uparrow}\rangle  + \langle n_{A\downarrow}\rangle  - \langle n_{B\uparrow}\rangle -\langle n_{B\downarrow}\rangle \neq 0.
\end{split}
\end{equation}
In Sec. \ref{strong-coupling}, we will consider FM(+OFM) state, and this is a state where $\Delta_{FM}$ is maximized first, and then $\Delta_{OFM}$ is maximized next. 

\begin{table}[!b]
\begin{ruledtabular}
\centering
\begin{tabular}{c|cccc}
``Angular momentum''& $(A, R)$ & $(A, L)$ & $(B, R)$ & $(B, L)$\\
\hline
$s$ & + & + & + & + \\
$p_{x}$ & + &- & + &-\\
$p_{y}$ & + & + &- &-\\
$d$ & +  & - & - & + 
\end{tabular}
\caption{Angular momentum and phase at each Fermi point.}
\label{angular}
\end{ruledtabular}
\end{table}

Next, we turn to the density wave for $r \neq r'$. In order to classify these phases, we label our particle-hole order parameters by the phases at each Fermi point and transferred momentum; there are 4 possible cases for interband ($q=k_A +k_B$) and intraband ($q= 2k_m$) order (Table \ref{angular}). $s$-wave has the same phases at all Fermi points. $p_{x}$-wave changes its sign under parity transformation, $R \leftrightarrow L$, and $p_{y}$-wave does under band exchange, $A \leftrightarrow B$. $d$-wave is odd under both transformations. Applying this classification, we find that $i=0$ and 1 are both $s$-wave, although the former is intraband type and the latter is interband type. We put `` ' '' for interband order to distinguish these two. $i =2$ and 3 is found to be interband and intraband $p_{y}$-wave accordingly. 

When a band is commensurate, we have another family of order parameters called ``bond" order (BOW), which is basically the density-wave slid from on-site to``on-bond", and is the same as dimerization. The only difference between site order and bond order is phase of the order parameter; $\Delta$ is real for on site order, and imaginary for bond order. We found that the energy gain is maximized when order parameter is real, indicating that always on-site order has lower energy. Therefore, we will ignore the bond orderings in the remainder of the paper.

To be complete, in two-leg ladder problems, $p'$CDW is more commonly called as orbital antiferromagnet (OAF)\cite{Chudzinski2008}, or staggered flux (SF) state\cite{Tsuchiizu2002}. $p_y$-density waves can be called orbital-density waves, but should be distinguished from PDW in Ref. \onlinecite{Tsuchiizu2002}, which is a bond-order.

We label the particle-particle channels for superconducting orders in the same manner (Table \ref{order}). In particular, when $k_A \neq k_B$, the order parameter with $i =0, 3$ has non-zero momentum, $q = k_A - k_B$; the order exhibits periodic structure in real space similar to FFLO (Fulde-Ferrell-Larkin-Ovchinnikov) state\cite{Fulde1964,Larkin1965}. However, in our case, there is no external field to split the spin up and down electrons. This possibility of FFLO state in multi-orbital system without external field was first pointed out by Padilha and Continentino\cite{Padilha2009}. The $d'$SS state often appears in two-leg ladder problems\cite{Schulz1996}.

\section{Method}
\label{method}
We employ the standard Hartree-Fock approximation, reducing the quartic part of the Hamiltonian to quadratic, $\hat{A} \hat{B} \cong \langle \hat{A} \rangle \hat{B} +  \langle \hat{B} \rangle \hat{A} -  \langle \hat{A} \rangle \langle \hat{B} \rangle$, where $ \langle  \hat{A} \rangle$ and $ \langle  \hat{B} \rangle$ are determined by minimizing the energy. These expectation values correspond to (quasi) long-ranged orders\footnote{In purely one-dimensional system, it is known that there is no long-range order corresponding to spontaneous symmetry breaking of continuous symmetry. In the mean field approximation, ordering tendency is overestimated leading to fictional long-range order.} induced either by forward scattering or by backscattering. Since it is not feasible to treat all the scattering processes in whole parameter space, we mainly investigated two regimes: the weak-coupling regime where back-scattering is dominant, and strong-coupling regime where both forward scattering and backscattering compete. This separation is motivated from the observation that Stoner's scenario of phase transition driven by forward scattering requires coupling to be larger than some critical value, although the backscattering always opens a gap even in the weak-coupling limit in 1D. Therefore, in the weak-coupling regime, we ignore forward scattering, and focus on back scattering. In the strong coupling regime, we first assume that forward scattering drives the system to some kind of density polarization, and then consider the effect of residual backscattering to these polarized states.

\subsection{Effect of backscattering}
In this subsection, we explain the treatment of backscattering since we can treat it in the same manner for both weak and strong coupling regime by using constant density of states (DoS). Although the detailed form of DoS is important to determinate the phase boundary between strong-coupling phases and weak-coupling phases, this approximation is justified within each regime: in the strong coupling regime, kinetic terms are less important than interactions; in the weak-coupling regime, electrons far away from Fermi energy is irrelevant. 

We first focus on a single band case. The quadratic Hamiltonian obtained by mean-field approximation can be diagonalized, and the system is gapped at Fermi energy, $\epsilon_{F}$. Using the energy $\epsilon$ measured from $\epsilon_{F}$, the new dispersion is found to be, 
\begin{equation}
\pm \sqrt{\epsilon^{2 }  + g^{2}\Delta^{2}}
\end{equation}
where $\Delta$ is order parameter in Eq. \eqref{ph} and Eq. \eqref{pp}, and $g$ is the corresponding coupling constant.  In our model, $g$ is expressed by some linear combinations of $U$ and $J$. Complete list of coupling constants expressed by $U$ and $J$ is given in Table \ref{coupling_const} of the Appendix. Note that there exist contributions from Umklapp processes at half-filling, and extra interband scattering when $k_A = k_B$. Under the assumption of constant DoS, the energy gain by this gap is given by 
\begin{equation}
\delta E = \nu \int_{-\Lambda}^{0} \left( \epsilon + \sqrt{\epsilon^{2 }  + g^{2}\Delta^{2}}\right)  d\epsilon - g\Delta^{2},
\label{gain}
\end{equation}
where $\nu$ is the density of states for single band, and $\Lambda$ is the cut-off or bottom of the band. The values of these parameters are different in the weak-coupling and strong-coupling regime, and we will explain it below. 

By maximizing the energy gain in terms of $\Delta$, we get the analytical solution to the gap equation,
\begin{equation}
\Delta = \frac{\Lambda}{g} \sinh ^{-1}\left( \frac{2}{\nu g} \right).
\label{sol}
\end{equation}
The stable condition for ordered phase is $g>0$. In order to obtain the phase diagram, we compare the energy of possible phases, and choose the order which gives smallest energy as the ground state. Thus, phase boundaries indicates 1st order transition from one minima to another without coexistent region. 

When multiple bands are involved, the calculation becomes more tedious. In our two band model, the important possibilities are: two bands have the same order independently by intraband scattering, or two bands have an order by interband scattering. For the former case, the Eq. \eqref{gain} remains the same where $\nu$ is density of states for each band, and total energy only depends the averaged density of states, $\nu_{intra} \equiv (\nu_{A} +\nu_{B})/2$. For the latter case, the dispersion becomes more complicated in general, though the final result depends only on single parameter, $\nu_{inter}^{-1} = (\nu_{A}^{-1}+\nu_{B}^{-1})/2$. Thus, there is an inequality between interband and intraband density of states,
\begin{equation}
\nu_{intra}  \geqslant \nu_{inter}.
\label{inq}
\end{equation} 
Therefore different density of states lead to suppression of interband scattering.

Next, we see the weak-coupling limit and the strong coupling limit of the above results.

\subsection{Weak coupling limit}
In the weak-coupling limit ($g \rightarrow 0$), Eq. \eqref{sol} is reduced to
\begin{equation}
\Delta \rightarrow \frac{2\Lambda}{g}e^{-\frac{2}{\nu g}},
\end{equation}
and the energy gain for single band by gap-opening is found to be
\begin{equation}
\delta E \simeq  \nu g^{2} \Delta^{2}.
\label{BCS}
\end{equation}
The density of states is fixed to be the value at Fermi energy, and the cut-off $\Lambda$ is taken to be small compared to band width, $4t$. The density of states at Fermi energy is connected to the Fermi velocity of the band, $v$, and it is given by $\nu = \left(2\pi v \right)^{-1}$. Therefore,  we see that the velocity difference suppresses interband processes from Eq. \eqref{inq}.

\subsection{Strong coupling limit}
In the strong coupling regime, along with the backscattering, we have forward scattering which induces static orders or density polarized states. We assume that this polarization is maximized in the strong coupling limit, and study the effect of back scattering on each polarized state. We will not consider partially polarized states, which might appear between non-polarized state and fully polarized state, because the possible intermediate phases are complicated and depend sensibly on the details.

As a constant density of states, we will use the averaged value for tight binding dispersion, $\nu = 1/(4t)$, since most of the electrons participate in density-wave formation in the strong coupling limit. The energy of each state consists of two parts: static energy, and energy reduced by backscattering. The former is simply given by the sum of kinetic terms, and static density-density interaction. As we take constant DoS to be $1/(4t)$, kinetic term becomes $2t n (n-1)$ with particle density $n$ in each band. The reduction of energy by density wave formation is obtained from Eq. \eqref{gain}. In particular, in the limit of $g \rightarrow \infty$, it becomes
\begin{equation}
\delta E \simeq \frac{1}{4}g \nu^{2}\Lambda^{2} -\frac{1}{2}\nu \Lambda^{2} + \cdots,
\label{strong-gain}
\end{equation}
with 
\begin{equation}
\Delta \simeq \nu \Lambda \left( \frac{1}{2} - \frac{1}{3\nu^{2 } g^{2}}\right) +\cdots.
\end{equation}
The backscattering tries to use all the electrons to form a density wave in the strong coupling limit, so the cut-off $\Lambda$ in Eq. \eqref{gain} is taken as the energy of band bottom.

\section{Weak-coupling phase diagrams}
\label{weak-coupling phase diagrams}
\begin{figure*}[!tb]
\centering
\subfigure[$k_A \neq k_B \neq \pi/2$]{
\includegraphics[scale=0.55]{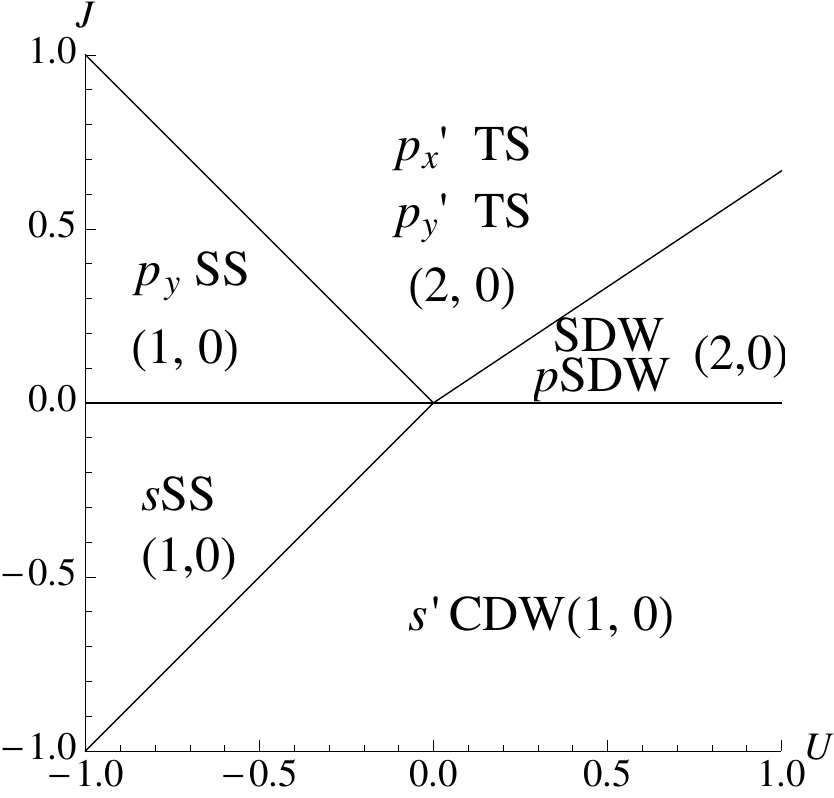}}
\subfigure[ $k_A \neq k_B = \pi/2$]{
\includegraphics[scale=0.55]{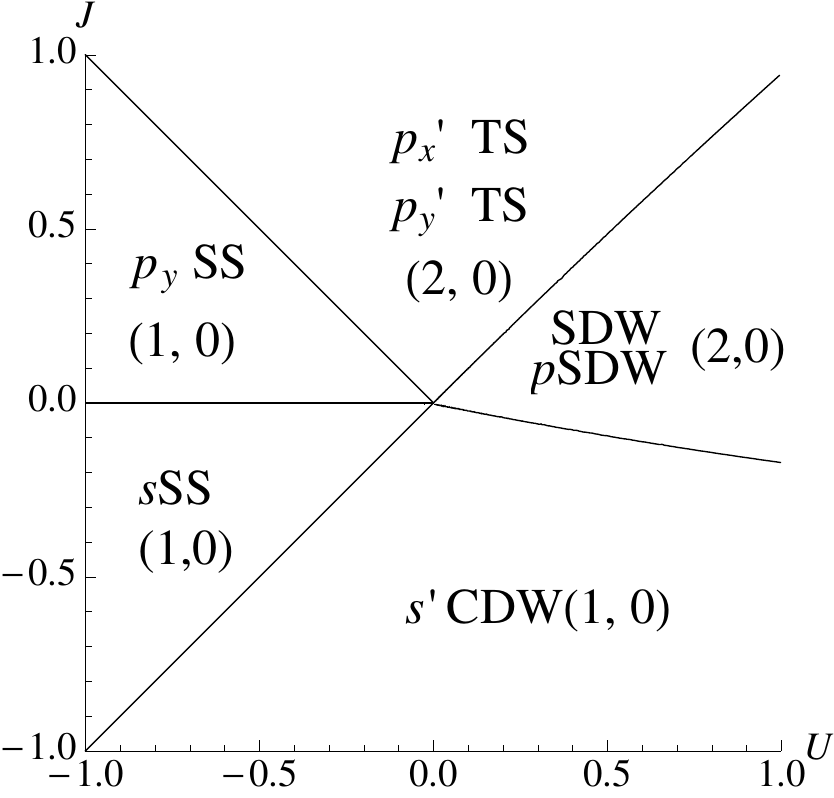}}
\subfigure[ $k_A = k_B \neq \pi/2$]{
\includegraphics[scale=0.55]{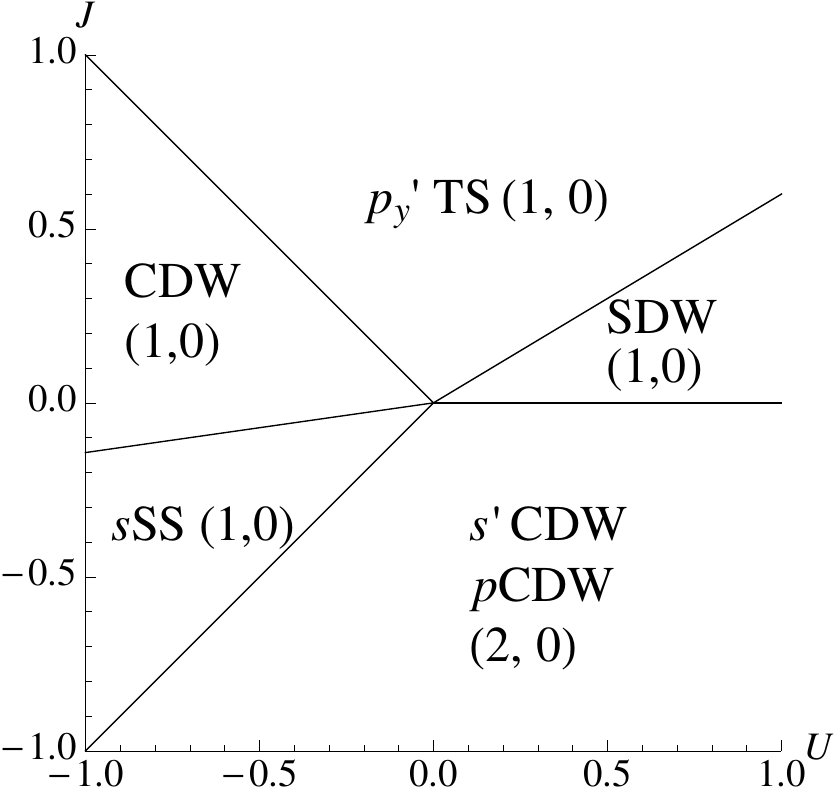}}
\caption{\label{incommensurate} Hartree-Fock phase diagrams for away from half-filling. Physically relevant region is $U >0$ and $U/2 > J > 0$. The number in the parenthesis indicates the number of massless modes in charge ($C$) and spin ($S$) sectors. We used $\nu = 1/4$ to get the phase diagram for $k_{A} \neq k_{B} = \pi/2$.}
\end{figure*}
Here we present Hartree-Fock phase diagrams in the weak-coupling regime. In order to obtain the phase diagrams, we compare the energy of possible phases, and choose the one with lowest energy as ground state. Along with the order parameter with (quasi) long-range correlation, these phases are characterized by the number of gapless excitations in charge and spin modes. We denote a system with $m$ massless charge modes and $n$ massless spin modes as $CmSn$\cite{Balents1996}. Without any interaction, the original Hamiltonian has 4 bands and this corresponds to $C2S2$.  

We first explain the three cases away from half-filling, and then see the phase diagrams for systems at half-filling. For all the cases, the physically relevant parameter region is $U\gg J>0$, although we investigated various parameter regions beyond this restriction.

\subsection{Away from half-filing} 
The weak-coupling phase diagrams where filling is away from half-filling are given in Fig. \ref{incommensurate} for three cases. We will look at each parameter region, and explain the dominant physics which governs the phase.
\subsubsection{$U \gg J >0$}
First we concentrate on upper right plane ($U, J >0$) since this is the physically relevant parameter region.  In the  small $J$ region, the  Coulomb repulsion $U$ dominates the physics, and as in the one orbital Hubbard model\cite{Overhauser1960} the ground state is a spin density wave. In the generic case of two incommensurate fermi wave vectors the phase of the spin density wave is not pinned between the two channels and there is a continuous family of solutions. In $k_A = k_B$ case, the two Fermi momenta are the same, and the relative phase mode is pinned down; there is no degeneracy here. 

In terms of number of massless modes, total charge mode and relative charge mode are both massless in degenerate SDW phase where both spin modes are massive. Thus, the degenerate SDW phase is expressed as $C2S0$. This represents two independent metallic spin-gapped chains of $C1S0$. Non-degenerate SDW phase in $k_A = k_B$ case has massive relative charge mode so it becomes $C1S0$.

\subsubsection{$J \gg |U| >0$}
The condition for Coulomb integrals to be positive ($U \gg J > 0$) means that this regime is unlikely to be relevant to real materials. One notable feature for large $J$ is that we have $p$-wave superconductivity, which is also observed in numerical calculations\cite{Sakamoto2002, Shirakawa2008}. This is different from the case of two-leg ladder systems, where purely repulsive Coulomb interaction leads to $d$-wave superconducting ground state of spin singlet \cite{Schulz1996}; the $p$-wave superconductivity is triggered by attractive interaction. We can understand this by looking at the limit of $J \rightarrow +\infty$, where the spin on the same site is fully polarized, but an orbital degeneracy remains. So the only on-site interaction with dynamical consequences is $(U-3J) n_{A\sigma}n_{B\sigma}$ in Eq. \eqref{int}. By employing the knowledge that negative-$U$ Hubbard model has spin singlet superconductivity as the ground state\cite{assa1994interacting}, we find that the analogous ground state of this limit is interband orbital singlet superconductivity with parallel spin, which is namely $p'_{y}$TS. The degeneracy of $p'_{y}$TS with $p'_{x}$TS arises from the absence of pinning effect between two SC as is the case for SDW and $p$SDW. When two Fermi momentum are not equal, these superconductivities show periodic modulation of order parameters in real space similar to that found in FFLO state.

At last, the degenerate $p'_{x}$TS and $p'_{y}$TS state is  $C2S0$, and non-degenerate $p'_{y}$TS for $k_{A} = k_{B}$ is $C1S0$.

\begin{figure*}[!Htb]
\centering
\subfigure[$k_A \neq k_B$]{
\includegraphics[scale=0.55]{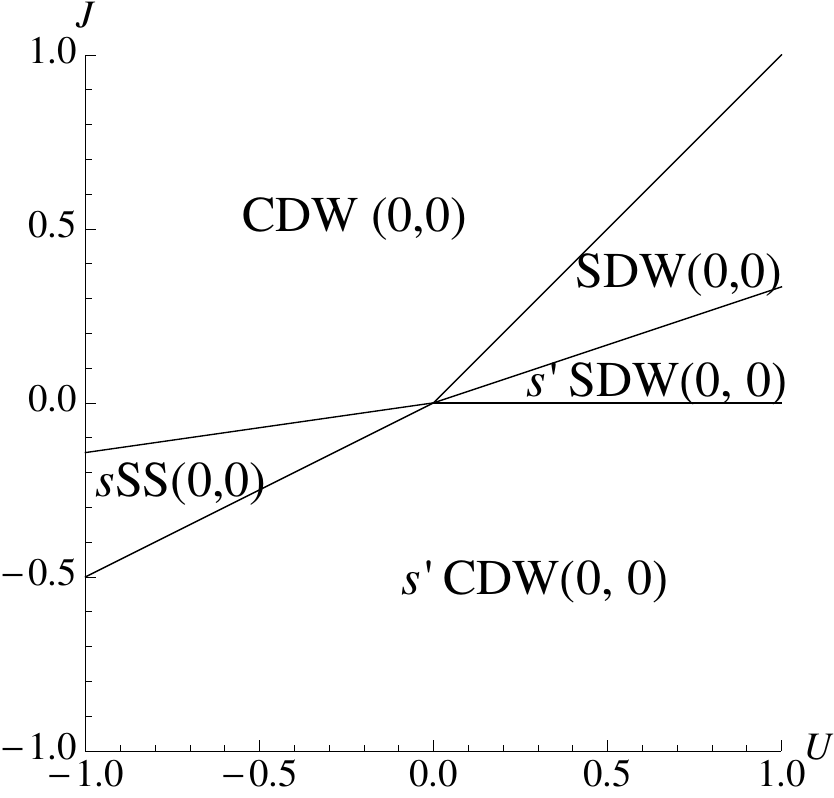}}
\subfigure[$k_A = k_B = \pi/2$ ]{
\includegraphics[scale=0.55]{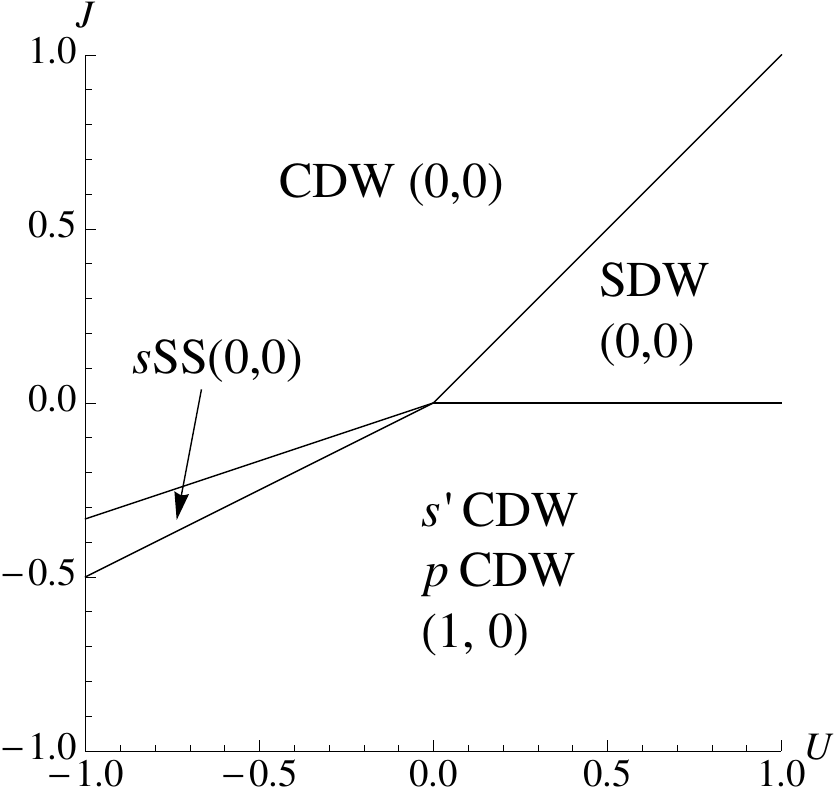}}
\caption{\label{commensurate} Hartree-Fock phase diagrams at half-filling. Physically relevant region is $U >0$ and $U/2 > J > 0$. The number in the parenthesis indicates the number of massless modes in charge ($C$) and spin sector ($S$). The $s$SS states should be read as S-Mott or S'-Mott state.}
\end{figure*}

\subsubsection{$-U \gg |J| >0$}
Next, we turn our attention to negative $U$ region. This regime is also not likely to be realized in transition metals. When $|U|$ is large enough, it is naturally expected that attractive $U<0$ gives some kind of superconductivity; indeed, we found $p_{y}$SS for positive $J$, and $s$SS for negative $J$ when two Fermi momenta are different. $p_{y}$SS is replaced to CDW when $k_{A} = k_{B}$.

To understand these phases, here we consider two-particle local eigenstates. There are ${}_4 C _2 = 6$ locally possible states. The spin triplet ($S =1$) states are
\begin{equation}
\begin{split}
&| S =1, S_z =1\rangle \equiv  c^{\dagger}_{A \uparrow} c^{\dagger}_{B \uparrow} | 0 \rangle\\
&| S =1, S_z =0\rangle \equiv   \frac{1}{\sqrt{2}}  \sigma^{x} _{s s'} c^{\dagger}_{A s} c^{\dagger}_{B s'} | 0 \rangle\\
&| S =1, S_z =-1 \rangle \equiv  c^{\dagger}_{A \downarrow} c^{\dagger}_{B \downarrow} | 0 \rangle\end{split}
\end{equation} 
The on-site energy is $E_{S=1}= U-3J$ . Among three spin triplet ($S =0$) states, $U(1)_\text{orbital}$ doublet states are
 \begin{equation}
 \begin{split}
| S=0, x \rangle &\equiv   \frac{1}{\sqrt{2}}  \sigma^{x} _{m m'} c^{\dagger}_{m \uparrow} c^{\dagger}_{m' \downarrow} | 0 \rangle\\
| S=0, z \rangle &\equiv  \frac{1}{\sqrt{2}}  \sigma^{z} _{m m'} c^{\dagger}_{m \uparrow} c^{\dagger}_{m' \downarrow} | 0 \rangle
\end{split}
\end{equation}
with $E_{S =0, -} = U-J$. The last piece is $U(1)_\text{orbital}$ singlet
\begin{equation}
| S=0,+ \rangle \equiv  \frac{1}{\sqrt{2}}  \delta_{m m'} c^{\dagger}_{m \uparrow} c^{\dagger}_{m' \downarrow} | 0 \rangle
\end{equation}
with $E_{S =0, +} = U+J$. This indicates that, for large negative $U$, interband superconductivity with $S =1$ is preferable for $J>0$, and intraband spin singlet superconductivity is preferable for $J<0$. The latter superconductivity is indeed $s$SS phase in negative $J$ region. On the other hand, the positive $J$ region doesn't match with $p_{y}$SS states in the phase diagram. This discrepancy is attributed to the different numbers of allowed scattering processes; when two Fermi momenta are different, the number of interband scattering process is fewer than that of intraband ones. Therefore, interband ordering is suppressed. For example, the following interband process is prohibited when $k_A \neq k_B$, 
\begin{equation}
n_{A\sigma}n_{B\bar{\sigma}} \simeq c_{AL\sigma}^{\dagger}c_{AR\sigma}c_{BR\bar{\sigma}}^{\dagger}c_{BL\bar{\sigma}} + \text{H.c.}
\end{equation}
although similar intraband process is allowed,
\begin{equation}
n_{m\sigma}n_{m\bar{\sigma}} \simeq c_{mL\sigma}^{\dagger}c_{mR\sigma}c_{mR\bar{\sigma}}^{\dagger}c_{mL\bar{\sigma}} + \text{H.c.}.
\end{equation}
Therefore, $| S=0, z \rangle$ is more suitable in positive $J$, and this corresponds to $p_{y}$SS. Of course, when $J$ becomes sufficiently strong, the energy gain by spin alignment becomes predominant, and the system exhibits spin-triplet superconductivity. Similarly, the CDW phase in upper left area for equivalent bands shows up since it is strongly enhanced  due to the ``nesting'' of $k_{A} = k_{B}$ although $p_{y}$SS is not affected.  

Here all the phases are non-degenerate, so only the total charge mode is massless, $C1S0$.

\subsubsection{$-J \gg |U| >0$}
Finally, the large negative $J$ region is again described by the $n_{A\sigma}n_{B\sigma}$ term in Eq. \eqref{int}, and ground state should be interband orbital density-wave, which corresponds to SDW of simple Hubbard model. So the possible candidates are either $s'$CDW, or $p$CDW. Taking into account the ordering of the fermionic operators in $n_{A\sigma}n_{B\sigma}$, we find that $s'$CDW has correct sign to be the ground state. For $k_A = k_B$ case, $s'$CDW and $p$CDW are degenerate as $U(1)_{\text{orbital}}$ symmetry requires.

\subsection{At half-filling}
The phase diagrams for half-filling cases are shown in Fig. \ref{commensurate}. At half-filling, the most of the argument of the general filling still apply, but we have to take Umklapp processes into consideration. Since Umklapp processes enhances only density-wave states, superconducting states which appear in negative $U$ region are now replaced by CDW as in $k_{A} = k_{B}$ case away from half-filling. An interesting new phase is $s'$SDW which is located between $s'$CDW and SDW for $k_{A}\neq k_{B}$ case. At this special filling, interband Umklapp process is enhanced, so $s'$SDW is dominant at small $J$. However, $s'$SDW is stable only when $U>J$, although SDW is stable for all $U, J>0$ region (See Table \ref{coupling_const} of the Appendix). Thus at large $J$, SDW is again dominant, and we obtain the above phase diagram. 

The $s$SS phase at half-filling should be read as S-Mott or S'-Mott state, which often appear in two-leg ladder problems\cite{Tsuchiizu2002}; at commensurate filling, we know that the system is insulating due to Umklapp process and not metallic. These Mott insulating states have similar order parameter as $s$SS except total charge mode when it is written in bosonic fields, and turn into $s$SS upon doping. 

Finally for most of the phases appearing at half-filling is completely gapped, $C0S0$, except a region where $s'$CDW and $p$CDW are degenerate. In this degenerate region, the orbital sector is massless, $C1S0$.

\subsection{Effect of velocity difference}
\begin{figure}[!tb]
\centering
\subfigure[$k_A \neq k_B \neq \pi/2$ at $n \neq 2$]{
\includegraphics[scale=0.45]{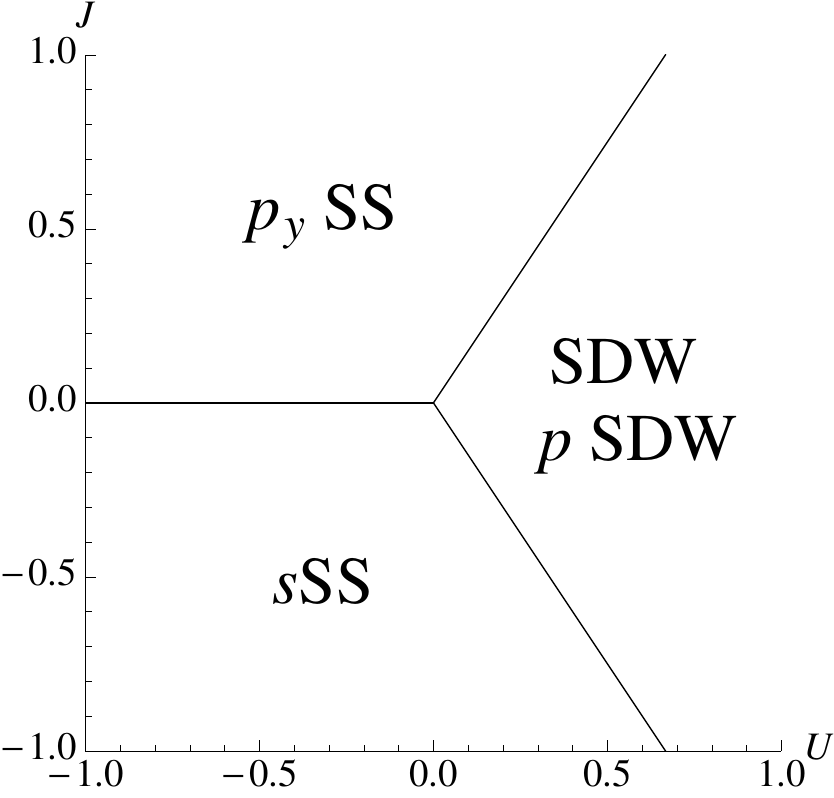}
\label{veldif1}}
\subfigure[$k_A \neq k_B$ at $n=2$ ]{
\includegraphics[scale=0.45]{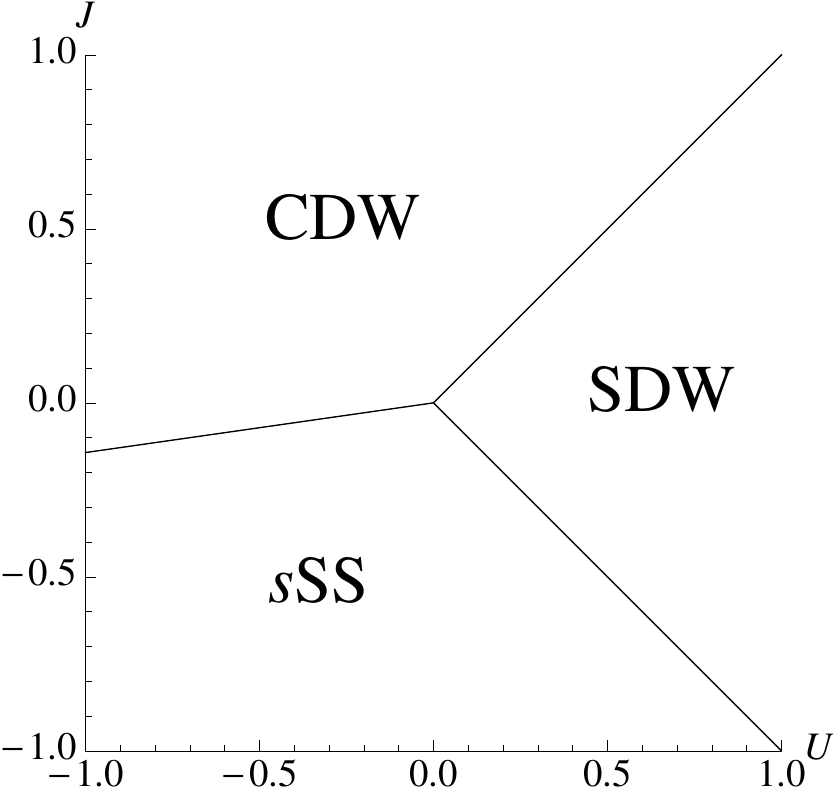}
\label{veldif2}}
\caption{\label{veldif} Phase diagrams for $v_{A}/v_{B} \sim40$. Physically relevant region is $U/2 > J > 0$. }
\end{figure}
As we pointed out in Sec. \ref{method}, the velocity difference suppresses interband scattering process, and intraband order becomes dominant. In our cases, the dominant phases appearing are $p_{y}$SS, CDW, SDW, $p$SDW and $s$SS. As the velocity difference gets larger from $v_{A}/v_{B} \sim1$, the phases governed by interband scattering are gradually excluded, and beyond $v_{A(B)} / v_{B(A)} \sim 40$, which corresponds to $\nu_{intra}/\nu_{inter} \sim10$, the whole phase diagram is covered by intraband type ordering (Fig. \ref{veldif}). The $k_{A} \neq k_{B} =\pi/2$ case looks like Fig. \ref{veldif1} when either $v_{A}/v_{B}$ or $v_{B}/v_{A}$ becomes large. When $k_{A} = k_{B}$, the phase diagrams are similar to Fig. \ref{veldif2} regardless of the filling.

\section{Strong-coupling phase diagrams}
\label{strong-coupling}
\begin{figure}[!htb]
\centering
\includegraphics[scale=0.4]{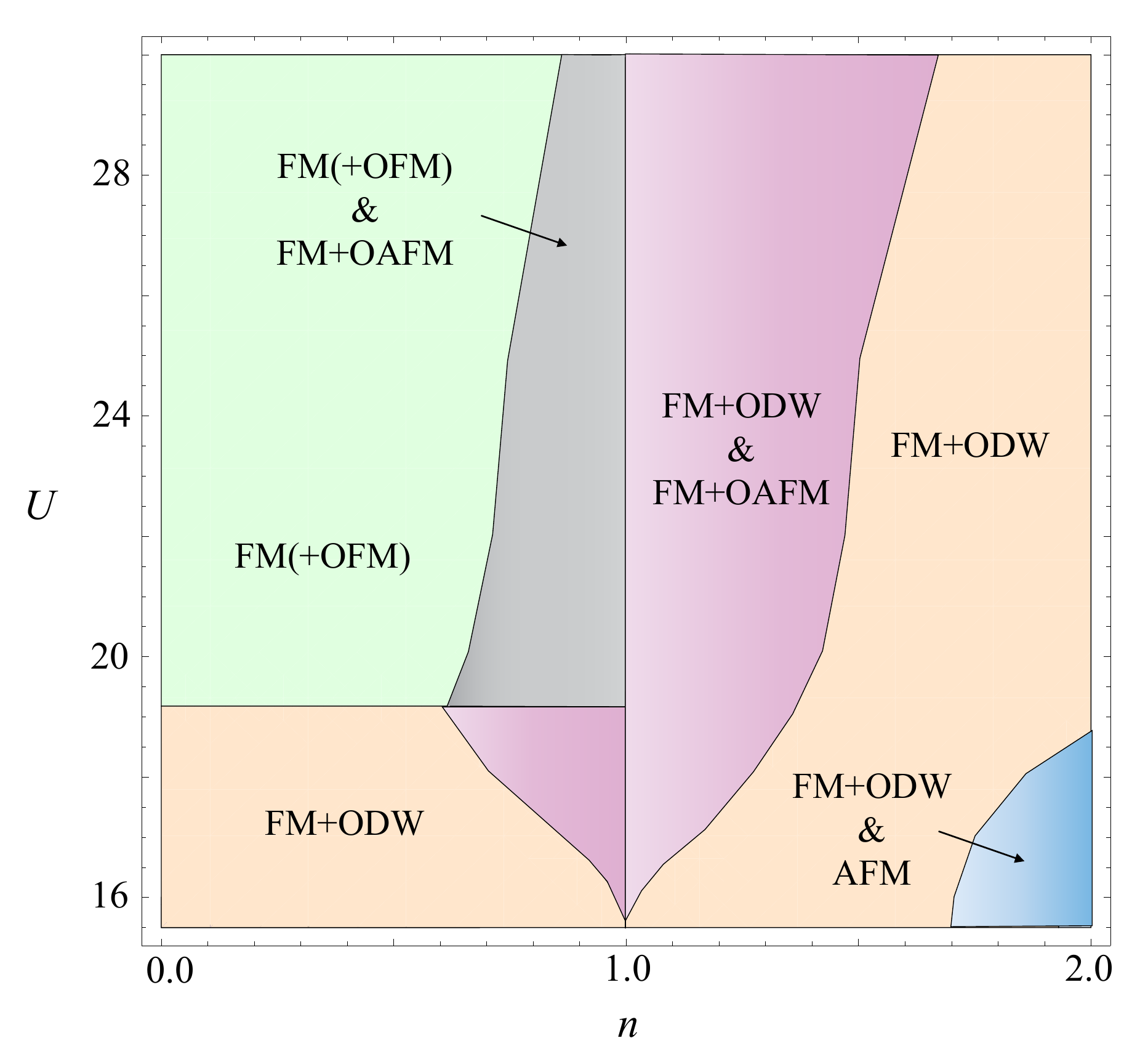}
\caption{Schematic phase diagram for the strong coupling limit, $U \gg J \gg t$ with $J/t=5$ including inhomogeneous phases. Green: FM(+OFM), Orange: FM+ODW, Gray: mixed state of FM(+OFM) and FM+OAFM, Purple: mixed state of FM+ODW and FM+OAFM, Blue: mixed state of FM+ODW and AFM.} 
\label{ps}
\end{figure}

We turn now to the phase diagrams in the strong coupling regime, $U\gg J \gg t$. Since we are mostly interested in high-spin state, we compared the energies of following 4 ferromagnetic states and 4 antiferromagnetic (or SDW) states: 
\begin{description}
\item[(1)]{FM, FM(+OFM), FM+ODW, FM+OAFM}
\item[(2)]{SDW, AFM, OFM+SDW, OFM+AFM,}
\end{description}
where AFM stands for antiferromagnetism, and OAFM is orbital-antiferromagnetism. In particular, we will distinguish two spin-density waves: AFM with $q=\pi$ and SDW with $q = 2k_F$. The two states are identical when $k_F=\pi/2$ but while the SDW state is driven by a fermi surface instability and is the only important state in the limit of weak coupling, the AFM state is stabilized by commensurability (Umklapp) effects and may exist for a range of carrier concentrations near the commensurate value. We similarly define orbital density wave (ODW) with wave vector $q=2k_F$, and OAFM with wave vector $q=\pi$.

Calculations very similar to those leading to Eq. \ref{gain} and Eq. \ref{sol} give the following results for $q=\pi$ orderings in a single band with $n <1/2$. The ground state energy is given by
\begin{multline}
E_{GS} =  \nu \int_{-2t}^{-2t(1-2n)}  \left( -\sqrt{\epsilon^{2}+ g'^{2 }\Delta^{2}}  \right) d\epsilon + g' \Delta^{2}\\
+ (\text{static interaction energy}),
\end{multline}
with $\nu = 1/(4t)$. We used $g' =2g$ to emphasize that the coupling constant is doubled at $q=\pi$ due to Umklapp process, while static energy from $q=0$ part is just with $g$. The kinetic energy for $n>1/2$ can be obtained by particle-hole symmetry. The solution for gap equation is found to be
\begin{equation}
\Delta = \frac{2t}{g' \sinh \left(\frac{2}{\nu g'} \right)} \sqrt{(2n-1)^2 +1 + 2(2n-1)\cosh\left(\frac{2}{\nu g'} \right)}.
\end{equation}
This goes to $\frac{n}{2}$ in the strong coupling limit. The solution exists only when the density is close to half-filling, $n_c < n < 1-n_c$, with
\begin{equation}
n_{c} = \frac{1}{2}\left( \sqrt{\cosh^{2}\left(\frac{2}{\nu g} \right) -1} - \cosh\left(\frac{2}{\nu g} \right)+1 \right).
\label{nc}
\end{equation}
This becomes 1/2 at $g \rightarrow 0$, and goes to 0 as $g \rightarrow \infty$. Thus, in the intermediate coupling, the density wave with $q=\pi$ is stable only around half-filling. 

For simplicity, we assumed two degenerate bands with constant density of states. Particle-hole symmetry allows us to investigate only $0<n<2$. Comparing the energies of the 8 states discussed above, we obtained ground state phase diagram, which is given in Fig. \ref{ps}. Below quarter-filling, FM(+OFM) -- where only single band is occupied -- is dominant with large $U$; this configuration does not cost any interaction energy. The transition between FM+ODW and FM(+OFM) below $n=1$ can be understood by Stoner's scenario where orbital sector becomes polarized above critical value, $U_{c}$. The precise behavior of the phase boundary as $n \rightarrow 0$ depends on details; for example, DoS of an isolated chain diverges at very small $n$ leading to the smaller value of critical interaction strength.

As the filling becomes closer to quarter-filling, FM+OAFM is found to be stable because it can use Umklapp processes to cancel static interaction energy, although this solution is unstable if too much holes or electrons are doped (See Eq. \eqref{nc}). As we plot the energies of these states, we found that phase separated state exists below $n=1$, which mixes FM+OAFM and FM(+OFM) for large $U$ and FM+OAFM and FM+ODW for small $U$. At exactly quarter-filling, the system is homogeneous FM+OAFM state.

Above quarter-filling, the FM+ODW is more stable than FM(+OFM) and FM+OAFM; the energy of FM+ODW in the strong coupling regime is roughly
\begin{equation}
E \simeq \frac{1}{8}(U-3J)n^{2},
\end{equation}
while the energy of FM(+OFM) and FM+OAFM is linear in $n$, $E = (U-3J)(n-1)$. Thus, FM+ODW is energetically preferable above $n\simeq 1.18$. Again the transition from FM+OAFM to FM+ODW is smeared by a phase separated state of these two.

Near half-filling, $n\simeq 2$, AFM state appears in weak $U$ regime by the same reason for FM+OAFM to appear around quarter-filling. However, the kinetic part of AFM does not cancel the static part completely, and the residual interaction makes this state unstable as $U$ gets larger. AFM state forms an inhomogeneous mixed state with FM+ODW below half-filling, and at half-filling, the system is totally occupied by AFM.


Now, we'd like to compare our mean-field phase diagram to the previously obtained results. At quarter filling ($n=1$), Kugel and Khomskii\cite{Kugel1972}, and Cyrot and Lyon-Caen \cite{Cyrot1975} found FM+OAFM as the ground state by strong-coupling expansion, and this is confirmed by numerical calculation\cite{Gill1987, Sakamoto2002}. This result can be understood as follows: when spins are totally aligned, Fermi momentum are doubled, and effectively the system is at half-filling. Then we may regard orbital index as pseudo spins, and the system exhibits pseudo-spin density wave.

Away from quarter-filling, the Umklapp process is killed so we expect OAFM is less dominant; indeed, Sakamoto \textit{et al.} \cite{Sakamoto2002} found FM+ODW with tight-binding DoS. They also found that adding far-neighbor hopping to get constant DoS replaces FM+ODW to paramagnetism ($S=0$) in $n<1$ though the system remains to be FM in $n>1$. This is because FM is induced by double-exchange mechanism for electron-doped case, but for hole-doped case, it is driven by purely one-dimensional ``spin-charge separation''\cite{Ogata1990}, which is fragile to perturbation of far-neighbor hopping. These observations do not contradict with our result above quarter-filling, though we have FM(+OFM) instead below quarter-filling. We think the FM(+OFM) state is actually more or less similar to the paramagnetic state without double occupancy in Ref. \onlinecite{Sakamoto2002}, since both configurations do not cost any interaction energy below quarter-filling. The mean-field treatment picks up FM(+OFM) among other configurations which do not have doubly occupied sites. On the other hand, the ferromagnetism in Ref. \onlinecite{Sakamoto2002} is induced by spin-charge separation, which is not a phenomenon captured by mean-field theory. Therefore, we conclude that the ferromagnetism of FM(+OFM) in Fig. \ref{ps} and the state seen in the numerical results of Ref. \onlinecite{Sakamoto2002} have different origins.

At half-filling, the system is claimed to be Haldane type where fully-polarized spin 1 on each site are antiferromagnetically coupled by exchange interaction. Slightly below half-filling, a phase separation between Haldane phase and FM+ODW was found\cite{Sakamoto2002}, which agrees with our results for small $U$.

\section{Summary}
\label{summary}
In this paper, we used mean-field theory to determine the phase diagrams of the two-band Hubbard model for a wide range of interactions including the intraband Coulomb repulsion $U $, interband Coulomb repulsion $U' (=U-2J)$, Hund coupling $J$, and pair-hopping $J$. For transition metal ions, we expect $U \gg J >0$.\\

First, we looked at the weak-coupling regime where back-scattering is dominant. For equal Fermi velocities, we observed five general features irrespective of band structure, each corresponds to the following parameter regions: 
\begin{description}
\item[(1) $U\gg J>0$] This parameter regime is relevant to real materials. In this region, various SDW orders (SDW, $s'$SDW, and $p$SDW) are most dominant similarly to simple Hubbard model. When $k_{A}\neq k_{B}$ with incommensurate filling, SDW and $p$SDW are degenerate since there is no phase pinning effect between independent SDWs in two orbitals due to the incommensurability. When $k_{A} +k_{B} = \pi$, the interband Umklapp process enhances $s'$SDW in small $J$ region, and becomes dominant. 
\item[(2) $J \gg |U| > 0$] The ground state is spin triplet superconductivity ($p'$TS) for incommensurate filling, and CDW at half-filling. The former is driven by the attractive interaction by large $J$ and the Cooper pair is formed by interband electrons. When $k_A \neq k_B$, their order parameters have non-zero momentum such as FFLO state\cite{Padilha2009} but without external field. The CDW at half-filing is induced by the strongly enhanced Umklapp process. 
\item[(3) $-U\gg -J>0$] This is again described by single-band Hubbard physics, and $s$SS order develops.
\item[(4) $-U\gg  J>0$] When filling is commensurate or two Fermi momentum are equivalent, these conditions allow additional scattering processes for CDW, and this becomes the ground state. On the other hand, for $k_{A} \neq k_{B}$ and incommensurate filling, intraband $p$SS is dominant due to positive $J$ and suppression of interband process.
\item[(5) $-J\gg |U| >0$] The ground state is interband CDW ($s'$CDW), which is an orbital analogue of SDW in single-band Hubbard model. When the system has $U(1)_{\text{orbital}}$ symmetry, $p$CDW is degenerate to $s'$CDW.
\end{description}

The velocity difference reduces interband process, and intraband ordering becomes dominant relatively. As we increase the velocity difference from $v_A/ v_B =1$, we observed that interband type ordering gradually expelled from the phase diagram, and above $v_A /v_B \sim 40$, the phase diagram is completely covered by intraband type ordering.

Second, we investigated the strong coupling regime of the model. We found ferromagnetism is almost always achieved, and various orbital orders are realized depending on density and interaction. Around quarter-filling, FM+OAFM  with $q=\pi$ modulation is stable. The region above quarter-filling is dominated by FM+ODW with $q=2k_F$, and there exists phase separation between these two phases. Below quarter-filing, FM+ODW is dominant for small $U$, and this is replaced by FM(+OFM) for larger $U$. The transition between these two phases by changing the magnitude of interaction can be understood by Stoner's scenario. As density gets larger, these states pass through phase separated regime and become FM+OAFM at quarter-filling . Close to the half-filling, there is an inhomogeneous mixed state of AFM and FM+ODW for small $U$.


For the Co/Cu system, it is located in $U \gg J > 0$ region if we neglect the hybridization between the Co wires and Cu surface. Although the real system has more than two orbitals, we may expect the following results about the Co/Cu systems from our calculations. First, if the system is in the weak-coupling regime, it exhibits SDW, $s'$SDW, or degenerate SDW-$p$SDW state depending on the band structure. In bosonization scheme, these phases are replaced by spin-gapped phases or quasi-long range order phase. In strong-coupling regime, we may have FM(+OFM), FM+ODW, or FM+OAFM. If the filling is very close to half-filling and the $U$ is not large, AFM is also possible. Of course the effect of hybridization or larger number of orbitals will introduce more complex physics into the system, but we will not pursue it here.

In a subsequent paper, we will present a result by renormalization and bosonization taking quantum fluctuation into consideration. The low energy effective Hamiltonian obtained after integrating out the high frequency modes could be different from the microscopic Hamiltonian we considered here, and different ground state is expected to appear. Especially several Mott insulating phases which are dual phases of density-wave should be investigated. These order parameters are expressed by non-local ``string'' operators, and are not considered here.

\begin{acknowledgements}
We thank R. Osgood, and N, Zaki for fruitful discussions and the data they shared with us. This work was supported by the Department of Energy Contract No. DE-FG 02-04-ER-46157.
\end{acknowledgements}

\begin{widetext}
\appendix 
\section{Coupling constants}
In this Appendix, we list up the coupling constants of the phases we considered in this paper (Table \ref{coupling_const}). The stable condition for the gap equation requires the coupling constant to be positive. For commensurate filling, Umklapp process doubles the number of possible scattering for density waves, though the superconducting order is not affected. When only one band is commensurate, the intraband scattering is enhanced for density-wave formation.  
\begin{table*}[!htb]
\centering
\begin{ruledtabular}
\begin{tabular}{l|ccccc}
& $k_{A} \neq k_{B} \neq \pi/2$ &  $k_{A} \neq k_{B} = \pi/2$ & $k_{A}=k_{B} \neq\pi/2$ & $k_{A}+k_{B} =\pi$ &$k_{A}=k_{B} =\pi/2$\\
\hline 
CDW 		&$-U$ 		&$-U, -2U$ 		&$-3U+5J$	&$-3U+5J$	&$-6U+10J$\\
SDW 		&$U$		&$U, 2U$		&$U+J$		&$U+J$		&$2U+2J$\\
$s'$CDW	 	&$U-5J$		&$U-5J$		&$U-5J$		&$2U-10J$	&$2U-10J$\\
$s'$SDW 		&$U-J$		&$U-J$		&$U-J$		&$2U-2J$		&$2U-2J$\\
$p'$CDW 		&$U-3J$		&$U-3J$		&$U-3J$		&$2U-6J$		&$2U-6J$\\
$p'$SDW	 	&$U-3J$		&$U-3J$		&$U-3J$		&$2U-6J$		&$2U-6J$\\
$p$CDW 		&$-U$		&$-U, -2U$		&$U-5J$		&$U-5J$		&$2U-10J$\\
$p$SDW 		&$U$		&$U, 2U$		&$U-J$		&$U-J$		&$2U-2J$\\
\hline
$d'$SS 		&$-U+J$		&$-U+J$		&$0$			&$-U+J$		&$0$\\
$p'_{y}$TS 	&$-U+3J$		&$-U+3J$		&$-2U+6J$	&$-U+3J$		&$-2U+6J$\\
$p_{y}$SS 	&$-2U+2J$	&$-2U+2J$	&$-2U+2J$	&$-2U+2J$	&$-2U+2J$\\
$d$TS 		&$0$			&$0$			&$0$			&$0$			&$0$\\
$s$SS 		&$-2U-2J$	&$-2U-2J$	&$-2U-2J$	&$-2U-2J$	&$-2U-2J$\\
$p_{x}$TS 	&$0$ 		&$0$			&$0$			&$0$			&$0$\\
$s'$SS 		&$-U-J$ 		&$-U-J$		&$-2U+2J$	&$-U+J$		&$-2U+2J$\\
$p'_{x}$TS 	&$-U+3J$		&$-U+3J$		&$0$			&$-U+3J$		&$0$\\
\end{tabular}
\end{ruledtabular}
\caption{Coupling constant for each phase}
\label{coupling_const}
\end{table*}
\end{widetext}
\newpage

\bibliographystyle{apsrev}
\bibliography{HF}
\end{document}